\documentclass[12pt,leqno]{amsart}
\usepackage{amssymb}
\usepackage{amsmath,amssymb}
\newtheorem{theorem}{Theorem}[section]

\newtheorem{proposition}{Proposition}[section]

\newcommand{\dddd}{\partial_z}
\newcommand{\ddddd}{\partial_{\overline{z}}}
\begin{document}
\theoremstyle{plain}
\newtheorem{MainThm}{Theorem}
\newtheorem{thm}{Theorem}[section]
\newtheorem{clry}[thm]{Corollary}
\newtheorem{prop}[thm]{Proposition}
\newtheorem{lem}[thm]{Lemma}
\newtheorem{deft}[thm]{Definition}
\newtheorem{hyp}{Assumption}
\newtheorem*{ThmLeU}{Theorem (J.~Lee, G.~Uhlmann)}

\theoremstyle{definition}
\newtheorem{rem}[thm]{Remark}
\newtheorem*{acknow}{Acknowledgments}
\numberwithin{equation}{section}
\newcommand{\eps}{{\varphi}repsilon}
\renewcommand{\d}{\partial}
\newcommand{\re}{\mathop{\rm Re} }
\newcommand{\im}{\mathop{\rm Im}}
\newcommand{\R}{\mathbf{R}}
\newcommand{\C}{\mathbf{C}}
\newcommand{\N}{\mathbf{N}}
\newcommand{\D}{C^{\infty}_0}
\renewcommand{\O}{\mathcal{O}}
\newcommand{\dbar}{\overline{\d}}
\newcommand{\supp}{\mathop{\rm supp}}
\newcommand{\abs}[1]{\lvert #1 \rvert}
\newcommand{\csubset}{\Subset}
\newcommand{\detg}{\lvert g \rvert}
\title[partial Dirichlet-to-Neumann map]
{Inverse boundary value problem for Schr\"odinger equation
in two dimensions}

\author[O. Imanuvilov]{Oleg Yu. Imanuvilov}
\address{Department of Mathematics, Colorado State
University, 101 Weber Building, Fort Collins CO, 80523 USA\\
e-mail: oleg@math.colostate.edu}

\author[M. Yamamoto]{Masahiro Yamamoto}
\address{Department of Mathematical Sciences,
University of Tokyo, Komaba, Meguro,
Tokyo 153, Japan \\e-mail:  myama@ms.u-tokyo.ac.jp}

\begin{abstract}
We relax the regularity condition on potentials of Schr\"odinger
equations in the uniqueness results in \cite{EB} and \cite{IY2}
for the inverse boundary value problem of determining a potential
by Dirichlet-to-Neumann map.
\end{abstract}

\maketitle \setcounter{tocdepth}{1} \setcounter{secnumdepth}{2}

Let $\Omega\subset \Bbb R^2$ be a bounded smooth domain with
$\partial\Omega=\cup_{j=0}^K\Sigma_j$ where $\Sigma_j$ are smooth
contours and $\Sigma_0$ is the external contour.  Let
$\nu=(\nu_1, \nu_2)$ be the unit outer normal to
$\partial\Omega$ and let $\frac{\partial}{\partial\nu} =
\nabla\cdot\nu$.

In this domain, we consider a Schr{\"o}dinger equation with
potential $q$:
\begin{equation}
(\Delta +q)u=0\quad\mbox{in}\,\,\Omega.
\end{equation}

Consider the full Cauchy data
\begin{equation}\label{popo}
\mathcal C_q= \left\{ \left(u,\frac{\partial u}{\partial \nu}\right)
\Biggl\vert_{\partial\Omega}; \thinspace\thinspace
(\Delta+q)u=0\quad\mbox{in}\,\,\Omega,\,\, u\in H^1(\Omega)
\right\}.
\end{equation}

By the inverse boundary value problem we mean an inverse problem of
determining a potential in (0.1) by the full Cauchy data. Such a
problem was formulated by Calder\'on \cite{C}. In the
two-dimensional case, we refer to Blasten \cite{EB}, Brown and
Uhlmann \cite{B-U}, Bukhgeim \cite{Bu}, Imanuvilov and Yamamoto
\cite{IY2}, Nachman \cite{N} and to Novikov  \cite{Nov} for the
stability estimate. In \cite{EB} the author proved that the full
Cauchy data uniquely determine the potential within piecewise
$W^1_p(\Omega)$ with $p>2$. The goal of this paper is to improve the
regularity assumptions on the potential $q$ in the inverse boundary
value problem and sharpen the results in \cite{EB} and \cite{IY2}.

As for the related problem of recovery of the two-dimensional
conductivity, Astala and P\"aiv\"arinta \cite{AP} established the
uniqueness result for $L^\infty(\Omega)$ conductivities, which
significantly improves the regularity assumption in \cite{N}.

If supports of Dirichlet data $f$ belong to a subboundary
$\widetilde{\Gamma}$ and observation of  the Neumann data restricted
on $\widetilde{\Gamma}$, we call all such pairs of Dirichlet and
Neumann data by partial Cauchy data. Under the assumption $q\in
C^{4+\alpha}(\overline\Omega)$, the uniqueness result  for the
partial Cauchy data was proved in Imanuvilov, Uhlmann and Yamamoto
\cite{IUY} for the case of  arbitrary subboundary
$\widetilde\Gamma$. Guillarmou and Tzou \cite{GZ} improved the
assumption on potentials up to $C^{2+\alpha}(\overline\Omega)$ with
partial Cauchy data. As for other uniqueness results for general
second order elliptic equation in the two dimensional case with
partial Cauchy data on arbitrary subboundary, we refer to
Imanuvilov, Uhlmann and Yamamoto \cite{IUY1}, Imanuvilov and
Yamamoto \cite{IY1}.

In dimensions $n \ge 3$, with the full Cauchy data, Sylvester and
Uhlmann \cite{SU} established the uniqueness of recovery of conductivity
in $C^2(\overline\Omega)$, and later the regularity assumptions were
relaxed up to $C^\frac 32(\overline\Omega)$ in P\"aiv\"arinta,
Panchenko and Uhlmann  \cite{PPU} and up to $W^\frac 32_p(\Omega)$
with $p>2n$ in Brown and Torres \cite{BT}.  A recent result by
Haberman and Tataru  \cite {Hab} establishes the uniqueness
for Lipschitz continuous conductivities.
For the case of partial Cauchy data, uniqueness
theorems were proved under assumption that a potential of the
Schr\"odinger equation belongs to $L^\infty(\Omega)$ (Bukhgeim and
Uhlmann \cite{BuU}, Kenig, Sj\"ostrand and Uhlmann \cite{KSU}).

Our main result is as follows
\begin{theorem}\label{popl}
Let  $q_1,q_2\in L^p(\Omega)$ with $p > 2$. If $\mathcal
C_{q_1}=\mathcal C_{q_2}$ then $q_1=q_2.$
\end{theorem}

Theorem 3.5 in \cite{Bu} announces the same result as our main
theorem, but the argument in lines 1-3 on p.27 clearly does not work
and therefore the proof in \cite{Bu} misses some details. The rest
part of the paper is devoted to the proof of the theorem \ref{popl}.
\\
\vspace{0.3cm}
\\
Throughout the paper, we use the following notations.

{\bf Notations.} Let $i=\sqrt{-1}$, $x=(x_1,x_2), x_1, x_2 \in {\Bbb
R}^1$, $z=x_1+ix_2$, $\overline{z}$ denote the complex conjugate of
$z \in \Bbb C$. We identify $x  \in {\Bbb R}^2$ with $z = x_1 +ix_2
\in {\Bbb C}$ and $\xi=(\xi_1,\xi_2)$ with $\zeta=\xi_1+i\xi_2$.
We set $\dddd = \frac 12(\partial_{x_1}-i\partial_{x_2})$, $\ddddd =
\frac12(\partial_{x_1}+i\partial_{x_2}).$ The tangential derivative
on the boundary is given by
$\partial_{\vec\tau}=\nu_2\frac{\partial}{\partial x_1}
-\nu_1\frac{\partial}{\partial x_2}$, where $\nu=(\nu_1, \nu_2)$ is
the unit outer normal to $\partial\Omega$.  By $\mathcal L(X,Y)$
we denote the space of linear continuous operators from a Banach
space $X$ into a Banach space $Y.$
Let $B(0,\delta)$ be a ball in ${\Bbb R}^2$ of radius $\delta$
centered at $0.$  \\
\vspace{0.3cm}
\\
Let us introduce the
operators:
$$
\partial_{\overline z}^{-1}g=-\frac 1\pi\int_\Omega
\frac{g(\xi_1,\xi_2)}{\zeta-z}d\xi_1d\xi_2,\quad
\partial_{ z}^{-1}g=-\frac 1\pi\int_\Omega
\frac{g(\xi_1,\xi_2)}{\overline\zeta-\overline z}d\xi_1d\xi_2.
$$

Then we have
\begin{proposition}\label{Proposition 3.0}
 {\bf A}) Let $1\le p\le 2$ and $ 1<\gamma<\frac{2p}{2-p}.$ Then
 $\partial_{\overline z}^{-1},\partial_{ z}^{-1}\in
\mathcal L(L^p( \Omega),L^\gamma(\Omega)).$
\newline
{\bf B})Let $1< p<\infty.$ Then  $\partial_{\overline z}^{-1},
\partial_{ z}^{-1}\in \mathcal L(L^p( \Omega),W^1_p(\Omega)).$
\end{proposition}
A) is proved on p.47 in \cite{VE} and B) can be verified
by using Theorem 1.32 (p.56) in \cite{VE}.

Consider a holomorphic function $\Phi(x,y)=(z-(y_1+iy_2))^2$ with
$y = y_1 + iy_2$.  We introduce two operators:
$$
\widetilde {\mathcal R}_{\tau}g = \frac 12 e^{\tau(\overline \Phi-\Phi)}
\partial_{ z}^{-1}(ge^{\tau(\Phi-\overline \Phi)}),\quad {\mathcal R}_{\tau}g
= \frac 12 e^{\tau(\Phi-\overline \Phi)}
\partial_{ \overline z}^{-1}(ge^{\tau(\overline \Phi-\overline \Phi)}).
$$

{\bf Proof.} Without loss of generality, we may assume that $\Omega$
can be taken as square $(-K,K)\times (-K,K)$ for sufficiently
large $K.$
Indeed $\Omega\subset\subset(-K,K)\times (-K,K)$
for sufficiently large $K > 0$.  We extend the potentials
$q_j$, $j=1,2$ by zero in $ (-K,K)\times (-K,K)\setminus
\overline{\Omega}$.
Consider the following Cauchy data:
\begin{equation}\label{popoty}
\widehat{\mathcal C}_q= \left\{ \left(u,\frac{\partial u}{\partial
\nu}\right) \biggl\vert_{\partial\Pi}; \thinspace\thinspace
(\Delta+q)u=0\quad\mbox{in}\,\,\Pi,\,\, u\in H^1(\Pi) \right\},
\end{equation}
where $\Pi=(-K,K)\times (-K,K).$
We claim that $\widehat{\mathcal C}_{q_1}=\widehat{\mathcal C}_{q_2}.$
Let $\left(u_1,\frac{\partial u_1}{\partial \nu}\right)
\in \widehat{\mathcal C}_{q_1}$ where $u_1$ is the
corresponding solution to the Schr\"odinger equation.
Consider the pair $(u_1,\frac{\partial u_1}{\partial
\nu})\vert_{\partial\Omega}.$
Since ${\mathcal C}_{q_1}={\mathcal C}_{q_2}$, there exists
a solution to the Schr\"odinger equation in
the domain $\Omega$ with the potential $q_2$ such that
$(u_1,\frac{\partial u_1}{\partial
\nu})\vert_{\partial\Omega}=(u_2,\frac{\partial u_2}{\partial
\nu})\vert_{\partial\Omega}.$
Then since $q_j\vert_{\Pi\setminus\Omega}=0$, we extend $u_2$ in
$\Pi\setminus\overline{\Omega}$ by setting $u_2=u_1.$
Then such a function $u_2$ satisfies the Schr\"odinger equation with the
potential $q_2$ in the domain $\Pi$ and $(u_1,\frac{\partial
u_1}{\partial \nu})\vert_{\partial\Pi}=(u_2,\frac{\partial
u_2}{\partial \nu})\vert_{\partial\Pi}.$

We set $U_0=1, U_1= \widetilde R_\tau(\frac
12(\partial^{-1}_{\overline z} q_1-\beta_1)), U_j=\widetilde
R_\tau(\frac 12\partial^{-1}_{\overline z} (q_1U_{j-1}))$ for all $j
\ge 2$. The choice of constant $\beta_2$ will be made later. We
construct a solution to the Schr\"odinger equation in the form
\begin{equation}\label{gavnuk}
 u_1=\sum_{j=0}^\infty e^{\tau \Phi}(-1)^jU_j.
\end{equation}

First we need to show that the infinite series is convergent in
$L^r(\Omega)$ with some $r > 2$.

\begin{proposition}\label{elka}
Let $u\in W^1_p(\Omega)$ for any $p>2.$
Then for any $\epsilon\in(0,1)$ there exists a constant
$C(\epsilon)$ such that
\begin{equation}\label{KJ}
\Vert \widetilde{\mathcal R}_\tau u\Vert_{L^2(\Omega)}\le
C(\epsilon)\Vert u\Vert_{W^1_p(\Omega)}/\tau^{1-\epsilon}.
\end{equation}
\end{proposition}
{\bf Proof.} Let $\rho\in C_0^\infty(B(0,1))$ and
$\rho\vert_{B(0,\frac 12)}=1.$ We set
$\rho_\tau=\rho(\root\of{\tau}\cdot).$ Since $\widetilde{\mathcal
R}_\tau u=\widetilde{\mathcal R}_\tau (\rho_\tau
u)+\widetilde{\mathcal R}_\tau ((1-\rho_\tau) u)$ for any positive
$\epsilon$, there exists $p_0(\epsilon)>1$ such that
 $\Vert e^{i\tau\psi}\rho_\tau u\Vert_{L^{p_0(\epsilon)}(\Omega)}\le
 C(\epsilon)\Vert u\Vert_{W^1_p(\Omega)}/\tau^{1-\epsilon}.$
Hence applying Proposition \ref{Proposition 3.0} and the Sobolev
embedding theorem, we have
\begin{equation}\label{J4}
\Vert\widetilde{\mathcal R}_\tau (\rho_\tau u)\Vert_{L^2(\Omega)}\le
C(\epsilon)\Vert u\Vert_{W^1_p(\Omega)}/\tau^{1-\epsilon}, \quad
\forall\epsilon\in (0,1).
\end{equation}
Observe that
\begin{eqnarray}\label{ippo}
\int_{\Omega} \frac{(1-\rho_\tau) u e^{\tau(\Phi-\overline\Phi)}}{\overline
z-\overline \zeta}d\xi=\int_{\Omega}\frac{(1-\rho_\tau) u\partial_\zeta
e^{\tau(\Phi-\overline\Phi)}}{\tau(\overline z-\overline
\zeta)\partial_\zeta\Phi}d\xi=\\
\int_{\partial\Omega}\frac{(\nu_1-i\nu_2)(1-\rho_\tau) u
e^{\tau(\Phi-\overline\Phi)}}{2\tau(\overline z-\overline
\zeta)\partial_\zeta\Phi}d\sigma- \int_{\Omega}\frac{1}{\tau(\overline
z-\overline \zeta)}\partial_\zeta \left(\frac{(1-\rho_\tau)
u}{\partial_\zeta\Phi}\right
)e^{\tau(\Phi-\overline\Phi)}d\xi
+\frac{(1-\rho_\tau)ue^{\tau(\Phi-\overline\Phi)}}{\tau
\partial_z\Phi}.\nonumber
\end{eqnarray}

Obviously, by the Sobolev embedding theorem, for any positive
$\epsilon$, there exists a constant $C(\epsilon)$ such that

\begin{equation}\label{J6}
\left\Vert \frac{(1-\rho_\tau)u}{\tau
\partial_z\Phi}\right\Vert_{L^2(\Omega)}\le C\Vert u\Vert_{W^1_p(\Omega)}/\tau^{1-\epsilon}.
\end{equation}
For the second term on the right hand side of (\ref{ippo}), we have
$$\left\vert\int_{\Omega}\frac{1}{\tau(\overline z-\overline \zeta)}
\partial_\zeta \left(\frac{(1-\rho_\tau) u}{\partial_\zeta\Phi}
\right)e^{\tau(\Phi-\overline\Phi)}d\xi\right\vert
\le\int_{\Omega}\left\vert\frac{1}{\tau^\frac 12(\overline z-\overline
\zeta)}\left(\frac{\partial_\zeta\rho(\root\of{\tau}x)
u}{\partial_\zeta\Phi}\right )\right\vert d\xi
$$
$$
+\int_{\Omega}\left\vert\frac{1}{\tau(\overline z-\overline \zeta)}
\left(\frac{(1-\rho_\tau)\partial_\zeta u}{\partial_\zeta\Phi}\right
)\right\vert d\xi + \int_{\Omega}\left\vert\frac{1}{2\tau(\overline
z-\overline \zeta)}\left(\frac{(1-\rho_\tau)
u}{(\partial_\zeta\Phi)^2}\right )\right\vert d\xi .
$$
The function $\frac{(1-\rho_\tau)\partial_\zeta
u}{\partial_\zeta\Phi}$ is uniformly bounded in $\tau$ in
$L^{p_1}(\Omega)$ for any $p_1\in (1,2).$ Applying Proposition
\ref{Proposition 3.0}, we have
\begin{equation}\label{J3}
\left\Vert \partial^{-1}_z \left(\frac{(1-\rho_\tau)\partial_z
u}{\tau\partial_z\Phi}\right )\right\Vert_{L^2(\Omega)} \le C\Vert
u\Vert_{W^1_p(\Omega)}/\tau.
\end{equation}
On the other hand for any $p_2\in (1,2)$
$$
\left\Vert\frac{\partial_z\rho(\root\of{\tau}\cdot)
u}{\partial_z\Phi}\right\Vert_{L^{p_2}(\Omega)}\le C\Vert
u\Vert_{C^0(\overline\Omega)} \left\Vert
\frac{1}{\partial_z\Phi}\right\Vert_{L^{p_2}(B(0,\frac{1}{\root\of{\tau}}))}\le
C \tau^{(2-p_2)/2p_2} \Vert u\Vert_{W^1_p(\Omega)}.
$$
Thanks to this inequality, applying Proposition \ref{Proposition
3.0} again we have:
\begin{equation}\label{J2}
\left \Vert\frac{1}{\tau^\frac 12}
\partial^{-1}_{z}\left(\frac{\partial_\zeta\rho(\root\of{\tau}\xi)
u}{\partial_\zeta\Phi}\right )\right\Vert_{L^2(\Omega)} \le C \Vert
u\Vert_{W^1_p(\Omega)}/\tau^{1-\epsilon}.
\end{equation}
For any $p_3>1$, we have
$$
\left\Vert
\frac{(1-\rho_\tau)u}{(\partial_\zeta\Phi)^2}\right\Vert_{L^{p_3}(\Omega)}\le
C\Vert u\Vert_{C^0(\overline\Omega)}\left\Vert
\frac{1}{(\partial_\zeta\Phi)^2}\right\Vert_{L^{p_3}(\Omega\setminus
B(0,\frac{1}{2\root\of{\tau}}))}\le C(p_3) \Vert
u\Vert_{W^1_p(\Omega)}\tau^{(2p_3-2)/2p_3}.
$$
Therefore
\begin{equation}
\left\Vert\partial^{-1}_z\left(\frac{(1-\rho_\tau)
u}{\tau(\partial_\zeta\Phi)^2}\right )\right\Vert_{L^2(\Omega)}\le
C\Vert u\Vert_{W^1_p(\Omega)}/\tau^{1-\epsilon}. \end{equation}

From the classical representation of the Cauchy integral,
we obtain
\begin{equation}\label{J1}
\left\Vert \int_{\partial\Omega}\frac{(\nu_1-i\nu_2)(1-\rho_\tau) u
e^{\tau(\Phi-\overline\Phi)}}{2\tau(\overline z-\overline
\zeta)\partial_\zeta\Phi}d\sigma\right\Vert_{L^2(\Omega)}\le C\Vert
u\Vert_{W^1_p(\Omega)}/\tau.
\end{equation}
 From
(\ref{J4})-(\ref{J1}) we have (\ref{KJ}). $\blacksquare$

We claim that the infinite series (\ref{gavnuk}) is convergent in
$L^r(\Omega)$ for all sufficiently large $\tau.$  Let $\tilde p\in
(2,p).$ By Proposition \ref{elka}, Proposition \ref{Proposition 3.0}
and H\"older's inequality yield the existence of a positive
$\delta(\tilde p)$ such that
\begin{equation}\label{sun}
\Vert \widetilde{\mathcal R}_\tau u\Vert_{L^\frac{p\tilde
p}{p-\tilde p}(\Omega)}\le \widehat C\Vert u\Vert_{W^1_{\tilde
p}(\Omega)}/\tau^\delta.
\end{equation}
 Using (\ref{sun}) we have
\begin{eqnarray}
\Vert U_{j}\Vert_{L^\frac{p\tilde p}{p-\tilde p}(\Omega)}\le
\frac{\hat C}{\tau^\delta}\Vert \frac 12\partial^{-1}_{\overline
z}(q_1U_{j-1})\Vert_{W^1_{\tilde p}(\Omega)}\le \frac{\widehat
C}{2\tau^\delta} \Vert
\partial^{-1}_{\overline z}\Vert_{\mathcal L(L^{\tilde p}
; W^1_{\tilde p}(\Omega))}\Vert q_1 U_{j-1}\Vert_{L^{\tilde
p}(\Omega)}\nonumber\\
\le\frac{\widehat C}{2\tau^\delta} \Vert
\partial^{-1}_{\overline z}\Vert_{\mathcal L(L^{\tilde p}
;W^1_{\tilde p}(\Omega))}\Vert q_1\Vert_{L^p(\Omega)}\Vert
U_{j-1}\Vert_{L^\frac{\tilde pp}{p-\tilde p}(\Omega)} \nonumber\\
\le\left(\frac{\widehat C \Vert \partial^{-1}_{\overline
z}\Vert_{\mathcal L(L^{\tilde p} ; W^1_{\tilde p}(\Omega))}\Vert
q_1\Vert_{L^p(\Omega)}}{2\tau^\delta}\right )^{j-1}\Vert
U_{1}\Vert_{L^\frac{p\tilde p}{p-\tilde p}(\Omega)} .
\end{eqnarray}
Therefore there exists $\tau_0$ such that for all $\tau>\tau_0$
$$
\Vert U_{j}\Vert_{L^\frac{p\tilde p}{p-\tilde p}(\Omega)}\le
\frac{1}{2^j} \Vert U_{1}\Vert_{L^\frac{p\tilde p}{p-\tilde
p}(\Omega)} \quad \forall j\ge 2. $$ Then the convergence of the
series is proved.

Since
\begin{eqnarray*}
&&(\Delta+q_1)(U_je^{\tau \Phi})=4\partial_{\overline z}\partial_z
(e^{\tau \Phi}\widetilde R_\tau(\frac 12\partial^{-1}_{\overline z}
(q_1U_{j-1})))+q_1U_je^{\tau \Phi}\\
=&& 2\partial_{\overline z}(e^{\tau\Phi}\frac
12\partial^{-1}_{\overline z} (q_1U_{j-1}))+q_1U_je^{\tau
\Phi}=q_1U_{j-1}e^{\tau \Phi}+q_1U_je^{\tau \Phi},
\end{eqnarray*}
the infinite series (\ref{gavnuk})
represents the solution to the Schr\"odinger equation. By
Proposition \ref{elka}, we have
\begin{equation}\label{kazi}
\left\Vert \sum_{j=2}^\infty (-1)^jU_j\right\Vert_{L^2(\Omega)}
= o\left(\frac 1\tau\right)\quad \mbox{as}\,\,\tau\rightarrow +\infty.
\end{equation}

Similarly we construct the complex geometric optics solution for the
Schr\"odinger equation with the potential $q_2$
\begin{equation}\label{liza}
v=\sum_{j=0}^\infty e^{-\tau \overline \Phi}(-1)^jV_j, \,\, V_0=1,
V_1={\mathcal R}_{-\tau} (\partial_{z}^{-1}q_2-\beta_2),
\,\,V_j={\mathcal R}_{-\tau} (\partial_{z}^{-1}(q_2V_{j-1})),
\end{equation}
where constant $\beta_2$ will be fixed later.

By Proposition \ref{elka} the following asymptotic formula holds
true:
\begin{equation}\label{kazi1}
\left\Vert \sum_{j=2}^\infty (-1)^jV_j\right\Vert_{L^2(\Omega)}
=o\left(\frac 1\tau\right)\quad \mbox{as}\,\,\tau\rightarrow +\infty.
\end{equation}

Since the Cauchy data (\ref{popo}) for potentials $q_1$ and $q_2$
are equal, there exists a solution $u_2$  to the Schr\"odinger
equation with the potential $q_2$ such that $u_1=u_2$ on
$\partial\Omega$ and $\frac{\partial u_1}{\partial\nu}
=\frac{\partial u_2}{\partial \nu}$ on $\partial\Omega.$ Setting
$u=u_1-u_2$ we obtain
\begin{equation}\label{pp}
(\Delta+q_2)u=(q_2-q_1)u_1\quad \mbox{in}\,\,\Omega, \quad
u\vert_{\partial\Omega}=\frac{\partial u}{\partial
\nu}\vert_{\partial\Omega}=0.
\end{equation}

Denote $q=q_1-q_2.$ Taking the scalar product of equation (\ref{pp})
and the function $v$ we have:
\begin{equation}
\int_\Omega q u_1vdx=0.
\end{equation}
By (\ref{gavnuk}), (\ref{kazi}), (\ref{kazi1}) and (\ref{liza}),
we have
\begin{eqnarray}\label{01}
0=\int_\Omega qu_1vdx=\int_\Omega q
e^{\tau(\Phi-\overline\Phi)}(1-U_1-V_1)dx
+ o\left(\frac{1}{\tau}\right)
= \int_\Omega
q e^{\tau(\overline\Phi-\Phi)}dx\\ +\frac 14\int_\Omega
(\partial^{-1}_{\overline
 z}q(\partial^{-1}_{
 z}q_1-\beta_1)+\partial^{-1}_{z} q(\partial^{-1}_{
 \overline z}q_2-\beta_2))e^{\tau(\overline\Phi-\Phi)}dx
+ o\left(\frac{1}{\tau}\right)\quad\mbox{as}\,\,\tau\rightarrow
 +\infty.\nonumber
\end{eqnarray}

Let $\{q_{j,\epsilon}\}_{\epsilon\in (0,1)}\subset
C^\infty_0(\Omega)$ be a sequences of functions such that
\begin{equation}\label{PO}
q_{j,\epsilon}\rightarrow q_j\quad \mbox{in}\,\,L^p(\Omega)\quad
\mbox{as}\,\,\epsilon\rightarrow +0, \,\,\,\,\forall j\in\{1,2\}.
\end{equation}
We set $q_\epsilon=q_{1,\epsilon}-q_{2,\epsilon},$ $g=\frac
14(\partial^{-1}_{\overline
 z}q(\partial^{-1}_{
 z}q_1-\beta_1)+\partial^{-1}_{z} q(\partial^{-1}_{
 \overline z}q_2-\beta_2)), g_\epsilon=\frac 14(\partial^{-1}_{\overline
 z}q_\epsilon(\partial^{-1}_{
 z}q_{1,\epsilon}-\beta_1)+\partial^{-1}_{z} q_\epsilon (\partial^{-1}_{
 \overline z}q_{2,\epsilon}-\beta_2)).$
By Proposition \ref{Proposition 3.0}, we see that
\begin{equation}\label{dirka}
g_\epsilon\rightarrow g\quad\mbox{in}\,\,C^0(\overline\Omega)
\quad\mbox{as}\,\,\epsilon\rightarrow +0.
 \end{equation}

We remind the following classical result of H\"ormander \cite{Ho}.
Consider the oscillatory integral operator:
$$
T_\tau f(x)=\int_\Omega e^{-\tau i\psi(x,y)}a(x,y)f(y)dy,
$$
where $\psi\in C^\infty(\Bbb R^2\times \Bbb R^2)$ and
$a\in C_0^\infty(\Bbb R^2\times \Bbb R^2).$
We introduce the following matrix
$$
H_\psi = (\partial^2_{x_iy_j}\psi).
$$
\begin{theorem}\label{opop} (\cite{Ho})
Suppose that $det H_\psi\ne 0$ on
$\mbox{supp}\, a.$
Then there exists a constant ${\widehat C} > 0$ such that
$$
\Vert T_\tau\Vert_{L^2\rightarrow L^2}\le \frac{\widehat C}{2\tau}.
$$
\end{theorem}

We set
$\psi(x,y)=2(x_1-y_1)(x_2-y_2).$  Then
$$
H_\psi(x,y)=\left (\begin{matrix} 0 & -2\\ -2 & 0\end{matrix}\right
)
$$
and $det H_\psi(x,y)=-4.$  Then the condition in Theorem \ref{opop}
holds true.

We set $a(x,y)=\chi(x)\chi(y)$ where $\chi\in C_0^\infty(\Bbb R^n)$
and $\chi\vert_\Omega\equiv 1.$ Then, by  Theorem \ref{opop},  there
exists a constant $C$ independent of $\tau$ such that
\begin{equation}
\Vert T_\tau\Vert_{L^2\rightarrow L^2}+\Vert
T_{-\tau}\Vert_{L^2\rightarrow L^2}\le C/\tau.
\end{equation}
Setting $f=(q-q_{ \epsilon})\chi_\Omega$, we have
\begin{equation}
\Vert T_\tau (q-q_\epsilon)\Vert_{L^2(\Omega)}+\Vert
T_{-\tau}(g-g_\epsilon)\Vert_{L^2(\Omega)}\le C(\epsilon)/\tau,
\quad C(\epsilon)\rightarrow 0\quad\mbox{as}\quad \epsilon
\rightarrow +0.
\end{equation}

Hence $\mbox{mes}\{x\in\Omega : \vert(T_\tau
(q-q_\epsilon))(x)\vert\ge C(\epsilon)\}\le 1/\tau^2. $
By the stationary phase argument (e.g., \cite{E}), we have
\begin{equation}\label{masa}
\int_\Omega q_\epsilon e^{\tau(
\Phi(z,y)-\overline\Phi(z,y))}dx=\frac{2\pi q_\epsilon(y)} {{\tau}
}+\mathcal C(y,\tau), \overline{\lim}_{\tau\rightarrow
+\infty}\frac{\sup_{y\in \Omega} \vert\mathcal
C(y,\tau)\vert}{\tau}
\end{equation}
$$
= 0\quad\mbox{as}\,\,\tau\rightarrow +\infty.
$$
Suppose that the function $q$ is not identically equal to zero.
Then there exists a positive number $\alpha$ such that $\mbox{mes}
\{x\in\Omega\vert\vert q(x)\vert\ge \alpha\}=\delta>0.$
 For any
$\epsilon\in (0,\frac 12)$ there exists $\tau_0$ such that
\begin{equation}\label{Z}\mbox{mes}\{x\in\Omega : \vert(T_\tau
(q-q_\epsilon))(x)\vert\le
C(\epsilon)\}\ge\mbox{mes}(\Omega)-\delta/9, \quad \forall
\tau\ge\tau_0.
\end{equation}

By (\ref{PO}) and Egorov's theorem, there exists a set $\mathcal
O\subset \{x\in\Omega\vert\vert q(x)\vert\ge \alpha\}$ such that
\begin{equation}\label{ZZ}\mbox{mes }
\mathcal O=\delta/5\,\,\mbox{ and} \,\,\sup_{x\in \mathcal O}\vert
(q-q_\epsilon)(x)\vert+\sup_{x\in \mathcal O}\vert
(g-g_\epsilon)(x)\vert\rightarrow 0 \,\,\mbox{as}\,\,
\epsilon\rightarrow +0.
\end{equation}
Then there exists $\epsilon_0>0$ such that
\begin{equation}\label{vova}
sup_{x\in \mathcal O}\vert (q-q_\epsilon)(x)\vert+sup_{x\in \mathcal
O}\vert (g-g_\epsilon)(x)\vert\le \frac{\alpha}{10},\quad\forall
\epsilon>\epsilon_0.
\end{equation}
Increasing $\epsilon_0$ if this is necessary, we may assume that
\begin{equation}\label{000}
C(\epsilon)\le \alpha/10, \quad\forall \epsilon>\epsilon_0.
\end{equation} Now let us fix $\epsilon$ and $\tau_0$ such that
(\ref{Z}) and (\ref{000}) hold true.
It follows from (\ref{ZZ}) and (\ref{Z}) that
$\mbox{mes}( \mathcal O\cap \{ x\in\Omega :\vert(T_\tau
(q-q_\epsilon))(x)\vert\le C(\epsilon)\})\ge \delta/10$ for all
sufficiently large $\tau.$
Hence there exists a sequence $\tau_k\rightarrow +\infty$ such that
we can choose a sequence $y(\tau_k)\in\mathcal O\cap
\{x\in\Omega : \vert (T_{\tau_k}(q-q_\epsilon))(x)\vert\le
C(\epsilon)\}$ satisfying $y(\tau_k)\rightarrow \widehat y.$

By the stationary phase argument and the fact that $\Omega$ is
square, setting $\beta_1=\partial^{-1}_{\overline z}q_{1,\epsilon}(\widehat
y)$ and $\beta_2=\partial^{-1}_{z}q_{2,\epsilon}(\widehat y)$, we have
\begin{equation}\label{02/}
\int_\Omega (\partial^{-1}_{\overline
 z}q_\epsilon(\partial^{-1}_{
 z}q_{1,\epsilon}-\partial^{-1}_{
 z}q_{1,\epsilon}(\widehat y))+\partial^{-1}_{z}
q_\epsilon (\partial^{-1}_{
 \overline z}q_{2,\epsilon}-\partial^{-1}_{
 \overline z}q_{2,\epsilon}(\widehat y)))
e^{\tau(\overline\Phi(z,y(\tau_j))-\Phi(z,y(\tau_j)))}dx
\end{equation}
$$
= o\left(\frac{1}{\tau_j}\right)
\quad\mbox{as}\,\,\tau_j\rightarrow
 +\infty.
$$
By (\ref{01}) and (\ref{02/}), we have
$$
0 = \left\vert\int_\Omega q
e^{\tau(\Phi(z,y(\tau_k))-\overline\Phi(z,y(\tau_k)))}dx+\int_\Omega g
e^{\tau(\overline\Phi(z,y(\tau_k))-\Phi(z,y(\tau_k)))}dx
+ o\left(\frac {1}{\tau_k}\right)\right\vert
$$
$$
= \left\vert\int_\Omega q_\epsilon
e^{\tau(\Phi(z,y(\tau_k))-\overline\Phi(z,y(\tau_k)))}dx\right\vert
$$
$$
-\left\vert\int_\Omega (q-q_\epsilon)
e^{\tau(\Phi(z,y(\tau_k))-\overline\Phi(z,y(\tau_k)))}dx\right\vert
- \left\vert\int_\Omega (g-g_\epsilon)
e^{\tau(\overline\Phi(z,y(\tau_k))-\Phi(z,y(\tau_k)))}dx\right\vert
- \left\vert o\left(\frac {1}{\tau_k}\right)\right\vert
$$
$$
\ge\frac{2\pi\vert
q_\epsilon(y(\tau_k))\vert} {{\tau}
}-\vert(T_\tau(q-q_\epsilon))(y(\tau_k))\vert
-\vert(T_{-\tau}(g-g_\epsilon))(y(\tau_k))\vert
- \left\vert o\left(\frac{1}{\tau_k}\right)\right\vert
\quad\mbox{as}\,\,\tau_k\rightarrow +\infty.
$$
By (\ref{000}) and (\ref{vova}), we obtain
$$
0\ge \frac{2\pi\vert q_\epsilon(y)\vert} {{\tau_k}
}-\frac{\alpha}{10\tau_k}-o(\frac {1}{\tau_k})\ge
\frac{9\alpha}{10\tau_k}-\frac{\alpha}{10\tau_k}
- o\left(\frac {1}{\tau_k}\right)
\quad\mbox{as}\,\,\tau_k\rightarrow +\infty.
$$
Then for sufficiently large $\tau_k$ we arrive at the contradiction.
$\square$

{\bf Acknowledgements}\\
Most part of the paper has been written during the stay of the first
named author in 2012 at Graduate School of Mathematical Sciences of
The University of Tokyo and he thanks the Global COE Program "The
Research and Training Center for New Development in Mathematics"
for support of the visit to The University of Tokyo.

\end{document}